\documentclass[10pt,letterpaper]{article}
\usepackage[top=0.85in,footskip=0.75in,marginparwidth=2in]{geometry}
\usepackage{subcaption}
\usepackage{amsmath}
\usepackage[utf8]{inputenc}

\usepackage{cite}

\usepackage{nameref,hyperref}

\usepackage{microtype}
\DisableLigatures[f]{encoding = *, family = * }

\raggedright
\usepackage{changepage}

\usepackage[aboveskip=1pt,labelfont=bf,labelsep=period,singlelinecheck=off]{caption}

\makeatletter
\renewcommand{\@biblabel}[1]{\quad#1.}
\makeatother

\usepackage{lastpage,fancyhdr,graphicx}
\usepackage{epstopdf}
\fancyheadoffset[L]{2.25in}
\fancyfootoffset[L]{2.25in}

\usepackage{color}

\definecolor{Gray}{gray}{.25}

\usepackage{graphicx}

\usepackage{sidecap}

\usepackage{wrapfig}
\usepackage[pscoord]{eso-pic}
\usepackage[fulladjust]{marginnote}
\reversemarginpar

\begin{document}
\newcommand{\chen}{changyao_chen_performance_2009}
\newcommand{\bowen}{cole_evanescent-field_2015}
\newcommand{\bachtold}{reserbat-plantey_electromechanical_2016}
\newcommand{\apell}{apell_high_2012}
\newcommand{\lee}{lee_self-heterodyne_2012}
\newcommand{\indexsens}{prasad_high_2016}
\newcommand{\gusynin}{gusynin_magneto-optical_2007}
\newcommand{\marsha}{parmar_dynamic_2015}
\newcommand{\cornellone}{zande_large-scale_2010}
\newcommand{\enhanced}{cai_enhanced_2016}
\newcommand{\dispsensopt}{castellanos-gomez_mechanics_2015}
\newcommand{\opticmod}{phare_graphene_2015}
\newcommand{\opticmodtwo}{sorianello_complex_2016}
\newcommand{\siliconmicro}{bogaerts_silicon_2012}
\newcommand{\chenthesis}{chen_graphene_2013}
\newcommand{\dispsenselec}{vadiraj_rao_high_2012}
\newcommand{\mechmode}{thomson_theory_1988}
\newcommand{\gnemstwo}{barton_high_2011}
\newcommand{\gnemsthree}{mathew_dynamical_2016}
\newcommand{\gnemsfour}{benameur_electromechanical_2015}
\newcommand{\gnemsfive}{smith_electromechanical_2013}
\newcommand{\gnemssix}{lee_electrically_2013}
\newcommand{\indexsenstwo}{gulik_refractive_2017}
\newcommand{\indexsensthree}{yang_silicon_2016}
\newcommand{\indexsensfour}{zhang_-chip_2016}
\newcommand{\dynamictwo}{dai_nonlinear_2012}
\newcommand{\dynamicthree}{reserbat-plantey_local_2012}
\newcommand{\dynamicfour}{kim_finite-size_2016}
\newcommand{\opticforce}{li_harnessing_2008}
\newcommand{\photothermal}{barton_photothermal_2012}
\newcommand{\pereira}{pereira_optical_2010}
\newcommand{\masssensone}{naik_towards_2009}
\newcommand{\masssenstwo}{chaste_nanomechanical_2012}
\newcommand{\forcesens}{weber_force_2016}
\newcommand{\chargesens}{bunch_electromechanical_2007}
\newcommand{\loss}{vlasov_losses_2004}
\newcommand{\anisotropic}{chang_experimental_2016}
\newcommand{\slope}{porzi_integrated_2017}
\vspace*{0.35in}

\begin{flushleft}
{\Large
\textbf\newline{On-Chip Optical Transduction Scheme for Graphene Nano-Electro-Mechanical Systems in Silicon-Photonic Platform}
}
\newline
\\
Aneesh Dash\textsuperscript{1},
S.K. Selvaraja\textsuperscript{1,**},
A.K. Naik\textsuperscript{1,*}\\

\bigskip
\bf{1} Centre for Nano Science and Engineering, Indian Institute of Science, Bangalore
\\
\bigskip
* Corresponding author: anaik@iisc.ac.in\\
** Corresponding author: shankarks@iisc.ac.in

\end{flushleft}

\section*{Abstract}
We present a scheme for on-chip optical transduction of strain and displacement of Graphene-based Nano-Electro-Mechanical Systems
(NEMS). A detailed numerical study on the feasibility of three silicon-photonic integrated circuit configurations is presented:
Mach-Zehnder Interferometer (MZI), micro-ring resonator and ring-loaded MZI. An index-sensing based technique using an MZI loaded
with a ring resonator with a moderate Q-factor of $2400$ can yield a sensitivity of $28 fm/\sqrt{Hz}$, and $6.5 \times 10^{-6}
\%/\sqrt{Hz}$ for displacement and strain respectively. Though any phase-sensitive integrated-photonic device could be used for
optical transduction, here we show that optimal sensitivity is achievable by combining resonance with phase-sensitivity.
\bigskip

Graphene-based NEMS devices have attracted a wide range of research-interest due to ultra-low mass-density and tunability of
electrical and mechanical properties~\cite{\chen,\gnemstwo,\gnemsthree,\gnemsfour,\gnemsfive,\gnemssix}. These devices hold promise
for ultra-low mass-sensing~\cite{\masssensone,\masssenstwo}, force-sensing~\cite{\forcesens}, charge-sensing~\cite{\chargesens}, and
to study non-linear dynamics. Electrical transduction schemes~\cite{\lee}are the predominant method for transducing the motion of
these devices. These methods rely on the modulation of conductance of graphene during vibration of the device. The transduction is
realized either by a direct measurement scheme~\cite{\chenthesis} at the frequency of vibration or by a frequency-mixed-down
technique~\cite{\chen}. However, the presence of a large background signal and small Signal-to-Noise Ratio (SNR) make the
implementation of this scheme challenging. Furthermore, the displacement sensitivities obtained with electrical transduction are
generally insufficient to observe the thermo-mechanical noise~\cite{\dispsenselec}.

Optical transduction techniques offer better displacement sensitivities (in $fm/\sqrt{Hz}$), large bandwidths, and signal with
minimal background~\cite{\dispsensopt}. However, most existing optical measurement schemes are based on free-space optics, where the
set-up is large and requires precise alignment of optical components~\cite{\cornellone,\bachtold}. Integrated optics has been used
for transduction of these devices, but the NEMS and optics are fabricated on different substrates and hence the challenge of
alignment persists~\cite{\bowen}. A completely integrated optical transduction scheme would help overcome such challenges.

Recent reports on graphene-based nano-photonic modulators explore the prospect of integrating graphene on a silicon-photonic platform
for electro-optic applications at Near-IR wavelengths~\cite{\opticmod,\opticmodtwo}. Similar integration of graphene-based NEMS
devices can allow for transduction by index-based sensing, which is a well-established field in silicon
photonics~\cite{\indexsens,\indexsenstwo,\indexsensthree,\indexsensfour}. Complex refractive index-based sensing allows us to capture
both dispersive and dissipative effects of graphene at Near-IR wavelengths~\cite{\opticmodtwo,\bowen}. Hence, we choose it over
purely reflection-based, refraction-based, or absorption-based sensing. Besides displacement, another important property to be
measured in these NEMS devices is the static strain, which sets the linear dynamic range of operation of these
devices~\cite{\marsha,\dynamictwo,\dynamicthree,\dynamicfour}. We show that index-based sensing in graphene offers a direct method
for transduction of strain.

We propose a transduction scheme with integration of the graphene NEMS device on a Silicon-photonic platform, operating around a
wavelength of $1550 nm$, with high sensitivity to strain and displacement. We have carried out extensive simulations using a finite
element model (FEM). We use the results for calculations of response to strain and displacement using MZI, ring resonator, and
ring-loaded MZI. We find that the response of the ring-loaded MZI is the best of the three. We then compute sensitivity of the
transduction scheme using ring-loaded MZI to strain and displacement.

\begin{figure}[htbp]
            \centering
            \includegraphics[width=0.44\columnwidth]{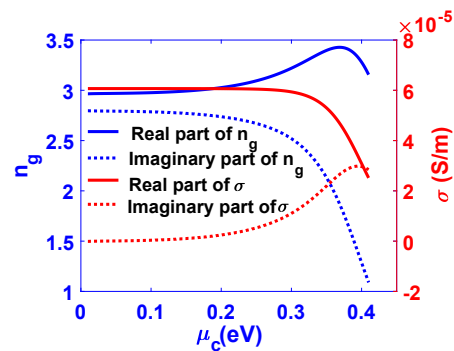}
            
\caption{\small  Refractive index $n_{g}$ (left axis) and Optical conductivity $\sigma$ (right axis) of graphene at $1550 nm$}             \label{f:index}   	    
\end{figure}

The in-plane refractive index of graphene and the geometry of the NEMS device over a waveguide-section determine the effective
refractive index, which governs the optical response of the device. The refractive index of graphene is given by $n_{g}(\omega) =
\sqrt{\epsilon_{g}(\omega)}$, where $\epsilon_{g}$ is the dielectric constant of graphene. For a monolayer graphene flake of
thickness $\Delta$ and an optical conductivity $\sigma$ as given in~\cite{\gusynin}, $\epsilon_{g}$ is given as~\cite{\apell}
\begin{equation}
\epsilon_{g}(\omega) = 1+\dfrac{j\sigma(\omega)}{\omega\epsilon_{0}\Delta} \, ,
\label{E:diel} 
\end{equation}
where $\omega$ is the optical frequency and $\epsilon_{0}$ is the permittivity of free space. Both conductivity and refractive index
are dependent on the chemical potential $\mu_{c}$. We use it to set the operating point for measuring strain and displacement. Since
the refractive index of graphene is anisotropic, we consider the out-of-plane index to be $1$~\cite{\anisotropic}.

Uniaxial strain $\epsilon$ in graphene leads to a change in the conductivity in the direction of strain as~\cite{\pereira}
\begin{equation}
		\sigma\approx\sigma_{0}(1-4\epsilon)\, ,
		\label{E:sig}
\end{equation}
where $\sigma_{0}$ is the optical conductivity at the operating point. We neglect the anisotropic variation of conductivity with
strain in the transverse direction, since we use a cross-sectional model in our analysis. The variation of $n_{g}$ and $\sigma$ with
$\mu_{c}$ in this range for $1550 nm$ wavelength is shown in Fig.~\ref{f:index}. We observe that at $\mu_{c}=\hbar\omega/2$, which is
$0.4 eV$ in Fig.~\ref{f:index}, there is sharp change in both $n_{g}$ and $\sigma$. This is due to the onset of Pauli-blocking of
inter-band transition in graphene. The value of $\delta n_{g}/\delta \sigma \approx 10^{4}$ at this point promises good sensitivity.
In addition, the real part of $n_{g}$ is maximum and hence the index contrast of graphene with the core and cladding of the waveguide
would be maximum at this point. Thus, the change in the effective refractive index would be dominated by the change in $n_{g}$ . We,
therefore, choose this as our operating point. We expect the sensitivity to strain and displacement to be maximum at this operating
point. The influence of strain on the conductivity and thus the refractive index of graphene can be utilized to ascertain the applied
static strain in the graphene-based NEMS devices. For calculating the response to strain, we use a flat graphene sheet over a
waveguide section to compute the effective indices for different values of strain.

\begin{figure}[htbp]
        \begin{subfigure}[b]{0.45\columnwidth}   
            \centering 
            \includegraphics[width=\textwidth]{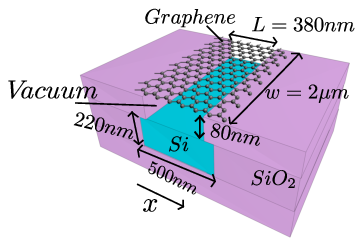}
          
            \caption[]
            {}    
            \label{f:cross}
        \end{subfigure}
        \begin{subfigure}[b]{0.45\columnwidth}
            \centering
			\includegraphics[width=\textwidth]{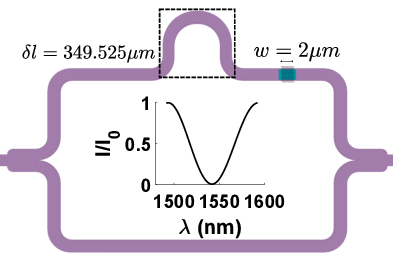}
			
			\caption[]
			{{}}
			\label{f:mzi_config}
         \end{subfigure}
        \vskip
		\baselineskip
		\vspace{-0.5 cm}
        \begin{subfigure}[b]{0.45\columnwidth}
            \centering
			\includegraphics[width=\textwidth]{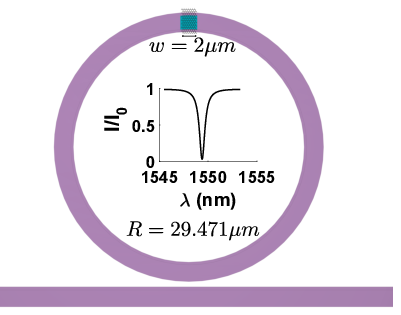}
			
			\caption[]
			{{}}
			\label{f:ring_config}
         \end{subfigure}
        \begin{subfigure}[b]{0.45\columnwidth}
            \centering
			\includegraphics[width=\textwidth]{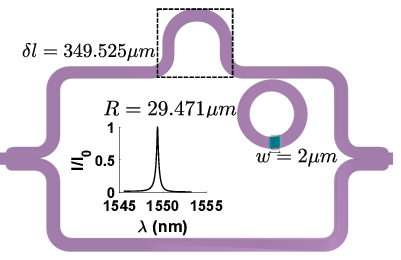}
			
			\caption[]
			{{\small }}
			\label{f:remzi_config}
         \end{subfigure}
        
\caption{\small (a) Cross-section of the wire waveguide, supporting quasi-TM mode, with suspended graphene (b) Schematic of MZI with
graphene: inset shows normalized intensity response with graphene, (c) Schematic of Ring resonator with graphene: inset shows
normalized transmission response with graphene, (d) Schematic of Ring-loaded MZI with graphene: inset shows the normalized intensity
response with graphene which has the smallest full-width at half-maximum (FWHM) of the three.}
        \label{f:config}
    \end{figure}

During vibration, the dynamic strain due to change in length of graphene will produce a proportional modulation in the index of
graphene. As graphene approaches and recedes from the waveguide during vibration, there will be a modulation in the effective
refractive index. Thus the vibration response for the fundamental mechanical mode with resonant frequency $\Omega$ is expected to
consist of a displacement component at frequency $\Omega$ and a strain component at frequency $2\Omega$. However, we find that this
$2\Omega$ component is not discernible in our vibration response for the dimensions of graphene considered. To calculate the
displacement response, we use the mode shape of the fundamental mode for a fixed-fixed beam, which is given as~\cite{\mechmode}
\begin{equation}	
\Phi(x) = \dfrac{cosh(bx)-cos(bx)-\sigma_{m}(sinh(bx)-sin(bx))}{\Phi_{m}} \, ,
\label{E:phi_mode} 
\end{equation}	
where $\Phi_{m}=1.6$, $b=\sqrt{22.4}/L$, $L$ being the length of the graphene beam, $x$ is position along the length of the beam as
shown in Fig.~\ref{f:cross}, and $\sigma_{m} = [cos(bL)-cosh(bL)]/[sin(bL)-sinh(bL)]$. The time-varying displacement at mechanical
resonance is given by $A\Phi(x)cos(\Omega t)$, where $A$ is the amplitude of vibration. We obtain the effective indices by using this
mode-shape of graphene, instead of a flat graphene sheet at different phases of vibration. This is because the interaction between
graphene and the optical mode in the waveguide varies as a function of $x$ and using a flat graphene sheet would overestimate the
interaction. We present the results for a moderate vibration amplitude of $100 pm$~\cite{\chen}.

The cross-section of the wire waveguide with suspended graphene, used for computing effective indices is shown in Fig.~\ref{f:cross}.
We use a waveguide cross-section of $500nm \times 220nm$ with oxide side-clad, supporting the quasi-TM mode. We have chosen the
quasi-TM mode over the quasi-TE mode because the former has greater modal overlap with the suspended graphene. This is because it has
large components of electric field both perpendicular to graphene and along the direction of propagation, due to the high
index-contrast of the Si core with the cladding, and thus shows a larger change in effective index. We use both the real ($n_{effg}$)
and imaginary ($k_{effg}$) parts of the effective index in our calculations to capture both electro-refractive and electro-absorptive
effects in graphene. Typical values obtained from our simulations are $n_{effg}=1.61$ and $k_{effg}=0.89 \times 10^{-3}$. Our
bandwidth of operation is from $1550 nm$ to $1570 nm$. Since we use a single wavelength in our calculations, we ignore
waveguide-dispersion in this band. For the length-scales we have considered, the loss in the waveguide-section with graphene and the
loss along the ring dominate over the propagation loss along Si waveguides, and hence it is neglected~\cite{\loss}.

The TM-waveguide cross-section with suspended graphene of length $w=2 \mu m$ forms a small part of an MZI, a micro-ring resonator,
and a ring-loaded MZI. The MZI has a phase-sensitive intensity response. The micro-ring resonator is a resonant device and hence has
a sharp phase transition at resonance. The MZI loaded with a ring leverages upon the phase response of the ring and the intensity
response of the MZI.

A schematic of the unbalanced MZI, with suspended graphene on the longer arm, is shown in Fig.~\ref{f:mzi_config}. For MZI without
graphene, destructive interference is observed at wavelength $\lambda_{int0}=\delta l/(m-1/2)$, where $\delta l$ is the optical path
difference between the two arms without graphene and $m$ is an integer corresponding to the number of the dark fringe. The wavelength
of destructive interference for an identical MZI with graphene $\lambda_{int}=[\delta l + (n_{effg}-n_{eff0})w]/(m-1/2)$, where
$n_{eff0}$ is the effective index in the absence of graphene. We have chosen $\delta l$ such that there is a dark fringe at $1550 nm$
in the absence of graphene. The transmitted intensity with graphene, assuming an amplitude splitting of $50\%$ in both arms for an
input intensity of $I_{0}$, is given as,
\begin{equation}	
I=I_{1}+I_{2}+2\sqrt{I_{1}I_{2}}cos(\Delta \phi) \, ,
\label{E:mzi_trans}
\end{equation}	
where $\Delta \phi = (2\pi /\lambda)[\delta l + (n_{effg}-n_{eff0})w]$, $n_{eff0}$ the effective index of the waveguide section
without graphene, $I_{1}=I_{0}/4.exp(-4\pi k_{effg}w/\lambda)$, and $I_{2}=I_{0}/4$. The normalized intensity response for $w=2 \mu
m$ is shown in the inset of Fig.~\ref{f:mzi_config}.

In a micro-ring resonator, the resonant electric field gets enhanced and hence, graphene interacts with a stronger electric field
~\cite{\enhanced}. The ring resonator with graphene is shown in Fig.~\ref{f:ring_config}. The resonant wavelength of the ring
resonator without graphene $\lambda_{res0} = n_{eff0}(2 \pi R)/n$, where $R$ is the radius of the ring and $n$ is the number of the
resonant mode. For a ring with self-coupling coefficient $r$ and amplitude coefficient $a$, the field-enhancement factor is given by
$FE = \sqrt{\sqrt{ra}/(1-ra)}$, as defined in~\cite{\enhanced,\siliconmicro}. The electric field, enhanced by a factor $FE$, induces
a surface current $J=\sigma FE$ in graphene. We account for this surface current by replacing $\sigma$ with $\sigma FE$ in
Eq.~\ref{E:diel} and re-computing the corresponding values of $n_{g}$ and $n_{effg}$ for the quasi-TM mode. The ring we have used has
$R=29.47 \mu m$ and $\lambda_{res0} = 1550 nm$. The attenuation along the ring is $\approx 10.6 dB/cm$, quality factor is $\approx
2400$, and $FE=1.9$. For suspended graphene covering a length $w$ along the ring, the modified resonant wavelength
$\lambda_{res}=[n_{eff0}(2\pi R-w)+n_{effg}(w)]/m$.

The transmission response with graphene at the through-port, for an input intensity $I_{0}$, is given by
\begin{equation}	
I=I_{0}\dfrac{a_{g}^2-2ra_{g}\cos(\phi)+r^2}{1-2ra_{g}\cos(\phi)+(ra_{g})^2} \, ,
\label{E:ring_trans}
\end{equation}		
where $\phi$ is the phase detuning from resonance and the effective amplitude coefficient with graphene is given by $a_{g}=a^{(1-w/(2
\pi R))}.exp(-2 \pi k_{effg}w/ \lambda)$. The normalized transmission response for $w=2 \mu m$ is shown in the inset of
Fig.~\ref{f:ring_config}.
 
The ring-loaded MZI with graphene is shown in Fig.~\ref{f:remzi_config}. The overall phase delay introduced by the ring resonator
with graphene is~\cite{\siliconmicro}
\begin{equation}	
\phi_{ring} = \pi + \theta + \arctan\big(\dfrac{r\sin\theta}{a_{g}-r\cos\theta}\big) + \arctan\big(\dfrac{r a_{g}\sin\theta}{1-r
a_{g}\cos\theta}\big) \,,
\label{E:phi_ring}
\end{equation}	
where $\theta = 2\pi \lambda_{res}/FSR(\lambda/\lambda_{res}-1)$ is the phase detuning from the resonance and
$FSR=\lambda_{res0}^2/(2\pi Rn_{effg})$ is the free-spectral range of the ring. The phase difference between the two arms of the
modified MZI is given as $\Delta \phi_{2}(\lambda) = 2\pi /\lambda[\delta l + \phi_{ring}(\lambda)]$.

In the absence of graphene, a bright fringe occurs at $1550 nm$. With suspended graphene over the ring, we obtain the normalized
intensity response for $w=2 \mu m$ using $I_{1}=I_{2}=I_{0}/4$ and $\Delta \phi_{2}$ instead of $\Delta \phi$ in
Eq.~\ref{E:mzi_trans}. The intensity profile is shown in the inset of Fig.~\ref{f:remzi_config}.

\begin{figure}[htbp]
        \begin{subfigure}[b]{0.44\columnwidth}
            \centering
            \includegraphics[width=\textwidth]{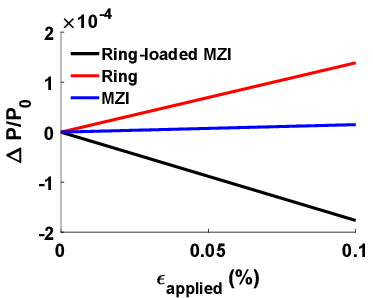}
            
            \caption[]
            {}    
            \label{f:strain}
        \end{subfigure}
        \begin{subfigure}[b]{0.44\columnwidth}
            \centering
			\includegraphics[width=\textwidth]{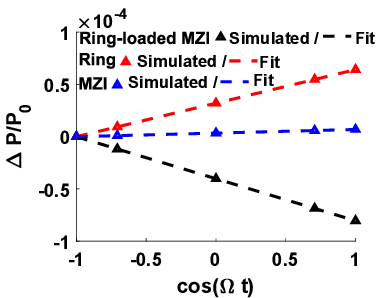}
			
			\caption[]
			{{}}
			\label{f:vibration}
         \end{subfigure}
        
\caption{\small (a) Power response to applied static strain: Ring-loaded MZI shows an overall change of $1.8 \times 10^{-4}$ which is
the best of the three, (b) Power response to vibration of amplitude $100 pm$: Fits are of the form $y=m_{1}cos(\Omega t)+c$
indicating linearity of the response to displacement. Ring-loaded MZI shows the best overall change of $0.8 \times 10^{-4}$. Changes
in $n_{effg}$ are of the order $10^{-6}$}
        \label{f:res_comb}
    \end{figure}

We choose the optimal probe wavelength that corresponds to the maximum gradient ($0.4 nm^{-1}$, $1.7 nm^{-1}$, and $8.8 nm^{-1}$) of
the intensity response of the MZI, ring resonator, and ring-loaded MZI respectively. We get the response to applied static strain,
shown in Fig.~\ref{f:strain}. $\epsilon_{applied}=0\%$ corresponds to the intrinsic strain. The ring-loaded MZI shows the best
overall change of $1.8 \times 10^{-4}$ for $0.1\%$ applied static strain. The changes in $\Delta P/P_{0}$ are of the order $10^{-4}$.
Optical probe powers can range from a few $nW$ to a few $\mu W$~\cite{\cornellone,\bachtold,\bowen}. For a probe power of $10 nW$,
the above-stated changes are detectable by commercially available InGaAs photo-detectors with noise-equivalent power (NEP) of the
order $fW/\sqrt{Hz}$ with less than $1 MHz$ measurement bandwidth.

The power response to a vibration amplitude of $100 pm$ is shown in Fig.~\ref{f:vibration}. The data-points correspond to five phases
during vibration. The ring-loaded MZI shows the best overall change with vibration ($0.8 \times 10^{-4}$). The responses show a
linear fit to $\cos(\Omega t)$. With a probing set-up identical to that considered for static strain, changes of this order are also
detectable with less than $1 MHz$ measurement bandwidth.
 
The noise in the responses depends on the input noise in the probe, quantified by the relative intensity noise (RIN), and the NEP of
the photo-detector. We consider RIN of the probe to be limited only by the photon shot noise. The RIN is lower for higher output
powers and the absolute output power in response to static strain and displacement depends on the input probe power. Thus, for
measuring extremely small changes in static strain and displacement with the ring-loaded MZI configuration, we use a relatively
higher input probe power of $10 \mu W$. The noise spectral density at this power and wavelength is $N_{s}=1.6 \times 10^{-12}
W/\sqrt{Hz}$, which is greater than the NEP ($=10^{-15} W/\sqrt{Hz}$) we have considered. So, the sensitivities are limited by the
source. We extract displacement and strain sensitivities of $398.3 fm/\sqrt{Hz}$ and $9.1 \times 10^{-5} \%/\sqrt{Hz}$ respectively
for $w=2\mu m$.

\begin{figure}[htbp]
        \centering
			\includegraphics[width=0.44\columnwidth]{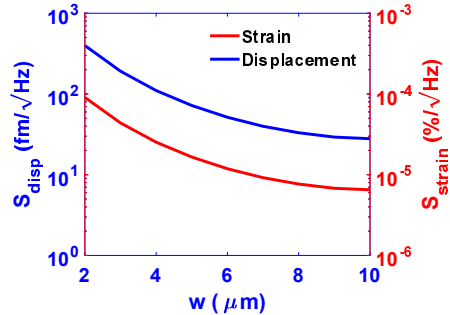}
			
\caption{\small Sensitivity of Ring-loaded MZI to Static strain and Displacement: Best sensitivities of $6.5 \times 10^{-6}
\%/\sqrt{Hz}$ and $28 fm/\sqrt{Hz}$ respectively at $w=10 \mu m$ }
            \label{f:sensitivity}
\end{figure}

As we increase $w$, the interaction length of graphene with the ring-loaded MZI increases leading to an improvement in the
sensitivities, as observed in Fig.~\ref{f:sensitivity}. For a gap of $80 nm$ under graphene and $L=380 nm$, it is possible to
fabricate suspended devices with $w$ of $10 \mu m$, beyond which there is a risk of collapse of graphene. For $w=10\mu m$, we compute
sensitivities of $28 fm/\sqrt{Hz}$ and $6.5 \times 10^{-6} \%/\sqrt{Hz}$ for displacement and strain respectively. These notable
sensitivities offer the ability to detect thermomechanical noise. The probe power of $10\mu W$ is at the input of the ring-loaded
MZI. In our configuration, the power in the waveguide section under graphene would be $\approx2.5\mu W$. For our quasi-TM mode, the
power at the surface of graphene would be lower by a factor of $100$. At these power levels, photothermal and optomechanical effects
due to the probe laser can be ignored~\cite{\bowen,\photothermal,\opticforce}. Hence this detection scheme has negligible
back-action.

We have shown that the ring-loaded MZI is the best of the three silicon-photonic devices considered for measuring displacement and
strain in graphene NEMS. It offers sensitivities of $28 fm/\sqrt{Hz}$ and $6.5 \times 10^{-6} \%/\sqrt{Hz}$ respectively. The
individual displacement and strain responses are linear. Even with a ring of moderate quality factor of $2400$, such high
sensitivities can be achieved. A moderate quality factor also allows for a large dynamic range of vibration amplitudes. Since this is
a completely integrated index-sensing based technique, there are no challenges of precise alignments and isolation of stray
reflections, as in other optical transduction schemes, while high displacement sensitivity is retained~\cite{\dispsensopt,\bowen}.
Further improvement in sensitivity can be achieved by reducing the gap between graphene and the Si waveguide or by using transduction
schemes with steeper intensity response~\cite{\slope}. The analysis here is general to any method of actuation, though each actuation
scheme would have its independent effect on the output and would have to be addressed separately. This technique can be used to study
thermomechanical noise and vibration-dynamics in graphene NEMS, with applications in mass-sensing, force-sensing, and charge-sensing.
Thus a ring-loaded MZI holds considerable potential for a highly sensitive, completely integrated silicon-photonic platform for
optical transduction in graphene NEMS.


\section*{Funding}
Council of Scientific and Industrial Research (CSIR), Government of India and Ministry of Electronics and Information Technology
(MeitY), Government of India.


\bibliographystyle{IEEEtran}
\bibliography{masssensone,masssenstwo,forcesens,chargesens,pereira,chen,gnemstwo,gnemsthree,gnemsfour,gnemsfive,gnemssix,lee,chenthesis,dispsenselec,dispsensopt,cornellone,bachtold,bowen,marsha,dynamictwo,dynamicthree,dynamicfour,indexsens,indexsenstwo,indexsensthree,indexsensfour,opticmod,opticmodtwo,gusynin,apell,anisotropic,mechmode,enhanced,siliconmicro,photothermal,opticforce,loss,slope}

\end{document}